\documentclass[conference]{IEEEtran}
\usepackage{amsmath,amssymb,amsfonts}
\usepackage{algorithmic}

 \usepackage{multirow}

\usepackage{graphicx}
\usepackage{lettrine}
\usepackage{textcomp}
\usepackage{xcolor}
\usepackage[
backend=biber,
style=ieee,
citestyle=numeric 
]{biblatex}
 \usepackage{booktabs}
 \usepackage{enumerate}

\begin{document}
%

\title{Hybrid Top-Down and Bottom-Up Approach for Residential Load Compositions and Percentages}

\author{\IEEEauthorblockN{Ahmed S. Alahmed}
\IEEEauthorblockA{ Electrical Engineering Department\\
King Fahd University of Petroleum and Minerals\\
Dhahran, Saudi Arabia\\
Email: alahmad@kfupm.edu.sa}
\and
\IEEEauthorblockN{Muhammed M. Almuhaini}
\IEEEauthorblockA{Electrical Engineering Department\\
King Fahd University of Petroleum and Minerals\\
Dhahran, Saudi Arabia\\
Email: muhaini@kfupm.edu.sa}
\thanks{This work was supported by the Deanship of Scientific Research (DSR) at King Fahd University of Petroleum and Minerals (KFUPM) through Project No. IN161043.}
\thanks{A variant of this work has been published in \cite{9542219}.}}

\IEEEoverridecommandlockouts
\maketitle
\IEEEpubidadjcol

\begin{abstract}

Load points are one of the most vital parts of power systems. Due to the new load forms and programs introduced in the demand side, the load serving entities (LSEs) no longer deal with lump loads, but rather with more dynamic, rational and price elastic loads. The high inter-temporal and behavioral variability of the load profile makes it almost impossible for utilities and system operators to expect the demand curve with the needed accuracy. A sound granularity of the load compositions and consumption percentages and patterns throughout the year is essential for avoiding energy losses, designing demand side management programs and ensuring proper adjustments of electricity rates. 
In this paper, a simplistic model that can be followed by system operators to initially understand the customers consumption pattern and the household load structure is proposed. A top-down approach is combined and matched with a detailed bottom-up one, to extract load compositions and percentages. Real and local load profiles integrated with household statistical data such as device time of use (ToU), number of device units per house and activities exercised in households are all included in the model. The main results of the paper show the load composition in residential demand and the percentage of such composition under seasonal-based scenarios.\\
\end{abstract}


\begin{IEEEkeywords}
Microgrids, controllable loads, demand-side management, load percentages, load composition, load modelling
\end{IEEEkeywords}

%

\IEEEpeerreviewmaketitle

\section{Introduction}

\lettrine{F}{or} a long time, electric utilities dealt with price-inelastic customers as technologies enabling load deferral and distributed generation were not widely deployed.
Nowadays, utilities are dealing with rational customers who optimize their consumption based on the announced prices. Under all circumstances, electric energy supply must always meet electric energy demand, and any unbalance in this equation may lead to brown-outs or more severely a blackout. LSEs worldwide are buying costly new generation whenever they speculate an increase in the demand during the peak hours, whereas this increase in demand can be easily addressed by understanding the load composition and applying data-driven DSM programs targeting the loads compositions with highest consumption percentages. The capricious load profile is no longer unanalyzable with the recent developments in Advanced Metering Infrastructure (AMI). 

\par
\begin{figure}[!htb] 
    \centering
    \includegraphics[trim=0 5 0 0,clip, width=\linewidth,height=5cm]{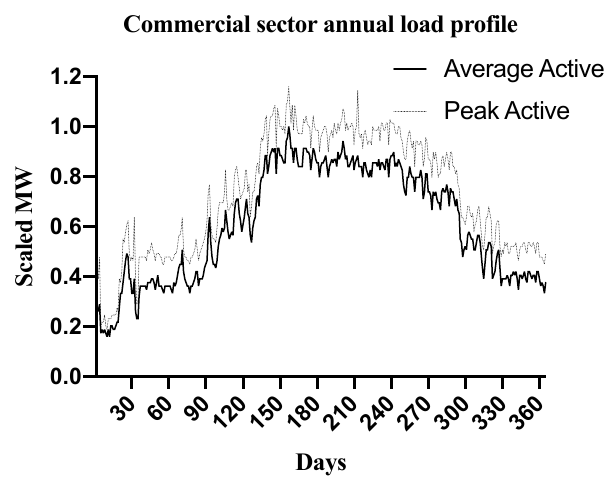} 
    \caption{Average and peak annual load profile.}
    \label{commdata} 
   \end{figure}

 System operators need to understand the load profile accurately before implementing demand side management (DSM) or imposing any time of use tariff on certain devices. Given a certain load profile , the controller must have information the components that contributed to this profile and the percentage of each load type (Fig. \ref{commdata}). Saudi Arabia has one of the fastest growing energy demand around the world with a population growth of 20\% between the years 2004 to 2010 \cite{general_authority_for_statistics}. For power systems such as the one in Saudi Arabia, 52\% of the energy sales come from the residential sector and thus system operators must understand the load composition properly in order to target specific home devices for energy conservation awareness or DSM programs.
 
\par Researchers tried to acquire a satisfactorily accurate determination of the household load composition using different techniques. There are approximately three known load modelling approaches : 1) component-based 2) measurement-based 3) A combination of the preceding two approaches (hybrid model). A comprehensive review about load modeling is given in \cite{arif_wang_wang_mather_bashualdo_zhao_2017}. 

\par Component based load modelling is a bottom-up method that aggregates the load information based on the composition of each load type and the characteristic of each load component. The component-based method was extensively studied in the literature \cite{lim_ji_ozdemir_singh_2001,dzafic2004a,4596557,gaikwad2016implementation,6303137,porsinger_janik_leonowicz_gono_2016}. For example, in \cite{lim_ji_ozdemir_singh_2001}, the authors used component-based load modelling to reduce the error between reactive design and the actual real value. Their model engages quantitative analysis and test of a cluster of loads on a real configuration of a substation. 

\par The measurement-based method is more widely used in microgrids (MG) given that  faster sampling and higher accuracy of data can be acquired from the distributed phasor measurement units (PMU). The  measurement-based approach, heavily relies on the data acquisition devices that are mounted at different locations in the system. The advantage of this method is the real-time accurately acquired data that does not require any estimation or variation of variables and thus it performs well in dynamic simulations. In 
\cite{papadopoulos_tzanidakis_papadopoulos_crolla_papagiannis_burt_2014, choi_chiang_li_li_chen_huang_lauby_2006, hou_xu_dong_2012, stojanovic_korunovic_milanovic_2008, maitra_gaikwad_zhang_ingram_mercado_woitt_2006, kontis_papadopoulos_chrysochos_papagiannis_2018, tasdighi_ghasemi_rahimi-kian_2014}, measurement-based method was used at different voltage levels. In \cite{papadopoulos_tzanidakis_papadopoulos_crolla_papagiannis_burt_2014}, the static and dynamic load model in grid-connected and islanded low-voltage (LV) MG was investigated.  The model was developed and then tested in a laboratory scale MG. In \cite{choi_chiang_li_li_chen_huang_lauby_2006}, online measurement data from Taiwan power system was gathered to derive, test and compare between different dynamic load models. The conducted numerical studies in the paper concluded that linear dynamic load models outperforms the nonlinear dynamic models when it comes to reactive power behavior modeling during disturbances. \cite{hou_xu_dong_2012} utilizes the measurement-based method to build up a complete load model at distribution level. The authors compared their model with a composite load model at transmission level and another model at generation level. They showed that their model performs better in transient conditions. References \cite{stojanovic_korunovic_milanovic_2008} and \cite{kontis_papadopoulos_chrysochos_papagiannis_2018} applied measurement-based dynamic load modelling using curve-fitting technique and vector-fitting technique ,respectively. \cite{tasdighi_ghasemi_rahimi-kian_2014}, discusses residential MG scheduling by utilizing smart meters to come up with a temperature dependent thermal load model. Sensitivity analysis is implemented to reflect the impact of the uncertainties contained in the model. A hybrid model was applied in \cite{li_yao_wang_zhu_yang_2015, milanovic_zali_2013,v_chakrabarti_srivastava_2015}. 

\par In \cite{li_yao_wang_zhu_yang_2015}, multiple data from single users is aggregated to generate the residential MG load profile. There are eight Major Electricity Consumption (MEC) events that when aggregated, a residential house load profile is acquired. The model parameters for each event are acquired using Ant Colony Optimization (ACO) algorithm. The load modeling method was then validated using a real MG in Ohio, USA. While in \cite{milanovic_zali_2013}, a dynamic equivalent Active Distribution Network Cell (ADNC) model was presented and examined. The model allows for more penetration of unconventional energy sources such as PV and wind in the distribution network. The performance of the model was tested using the modified IEEE nine bus system during different levels of disturbances. 

\par Although load modeling was appropriately studied in the literature, very few papers tackled exploring the load composition and percentages of real load data using simply the total load information and statistical data based on the location, ToU of the appliances, approximated number of units per house and load daily operation patterns. In this paper, a detailed load modeling will be implemented on real and local residential loads data from Saudi Arabia. Firstly, a top-down approach will be utilized to acquire annual, monthly and daily load profiles. Then, a bottom-up approach will be modeled on the residential load sector to identify the composition of loads and consumption percentage of each appliance to the whole consumption in summer and winter. 

\par In the following section the residential load structure and load composition modeling will be presented and the hybrid approach will be formulated and explained. The simplistic method to find load structures is discussed in section 3, and 
the results follow in section 4. Lastly, section 4 conclude the research with some remarks and outcomes of the study. 

\section{Residential Load Modeling}
The majority of energy sales in Saudi Arabia is attributed to residential costumers (Table \ref{tab:loaddata}). As a result, it is essential to investigate the load composition and percentages of household costumers.
Utilities carefully study the consumer behavior, which is mainly governed by the type of appliances, devices and machines the consumer is using. Each of these loads has a certain contribution to the residential load profile, which varies by seasons, weather, special events and prices. In order to conduct load controlling and hence demand-side management in a MG, a detailed identification of the load composition and its percentages must be determined. 

\par
\begin{figure}[!htb]
\centering
  \includegraphics[trim=0cm 4.5cm 0cm 5cm,clip,width=8cm,height=5cm]{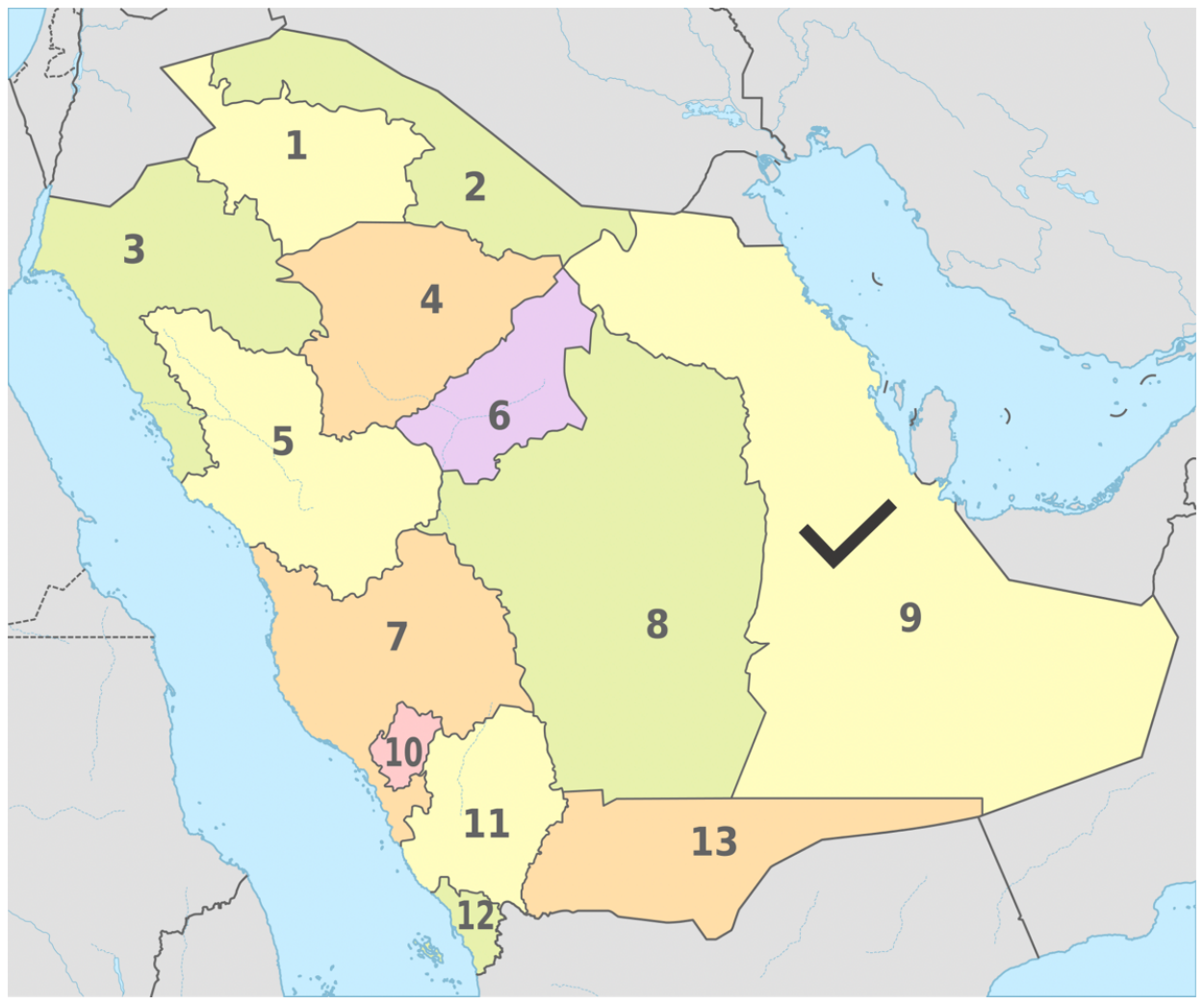}
  \caption{Eastern province in Saudi Arabia indicated with a tick mark.}
  \label{Eastern}
\end{figure}

 The location of data used in this paper is the Eastern province in Saudi Arabia, which is the largest province by area (Fig. \ref{Eastern}). The coordinates of the data location are 26.2361° N, 50.0393° E. The load data was acquired based on supervisory control and data acquisition (SCADA) measurements.

\begin{table}[!htb]
\centering
\caption{Energy Sales in Saudi Arabia Per Category \cite{EXRA}}
\label{tab:loaddata}
\resizebox{0.48\textwidth}{!}{%
\begin{tabular}{@{}cccc@{}}
\toprule
Classification & Ratings     & Energy Sales (GWh) & Energy sale (\%) \\ \midrule
Residential    & 0.4kV/0.22kV & 143,055            & 49.6\%           \\
Commercial     & 0.4kV/0.22kV & 48,252             & 16.72\%          \\
Industrial     & 13.8kV & 47,230             & 16.4\%           \\
Governmental   & 0.4kV/0.22kV & 38,422             & 13.3\%           \\
Others         & 13.8kV/0.4kV/0.22kV & 11,698             & 4.1\%            \\ \bottomrule
\end{tabular}}
\end{table}

\par
\begin{figure}[!htb]
\centering
  \includegraphics[width=0.8\linewidth,height=5cm]{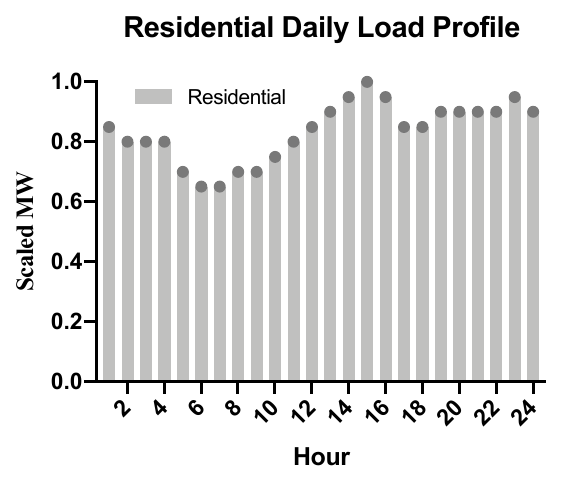}
  \caption{Residential daily load profile.}
  \label{dailyRES}
\end{figure}

\begin{figure}[!htb]
\centering
  \includegraphics[trim=0cm 0cm 0cm 1cm,clip,width=0.85\linewidth,height=5cm]{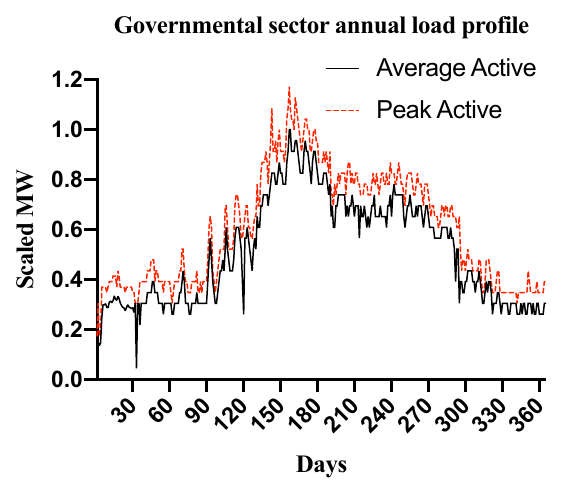}
  \caption{Residential annual load profile.}
  \label{ANNRES}
\end{figure}

 Using the measurement based method, which is a top-down approach, we acquire the residential annual and daily load profiles shown in Figs. \ref{dailyRES} \& \ref{ANNRES}. The residential daily load profile shows the peculiar habit of household consumption in Saudi Arabia, where the load peaks at around 3 P.M due to consumers coming out of work and schools. The consumption is minimum at 6 A.M that is when most customers are sleeping. Most of the manually operated loads (i.e., oven, vacuum cleaner, and TV) in this period are idle while the automatically and semi-automatically operated loads (i.e., refrigerator and air conditioner) are less affected by the factor of time.

\par The annual average and peak load profile are displayed in Fig. \ref{dailyRES} \& \ref{ANNRES}. All figures were scaled as per the following equation: 
\begin{equation}
    P_{s}(t)= \frac{P_{l}(t)}{P_{max}}
\end{equation}
Where, $\{P_{s} \in \mathbb{R}_+ \:|\:0< P_{s}<1$\} . $P_{max} \in \mathbb{R}_+$ is the daily maximum power, and $P_{l}(t)\in \mathbb{R}_+$ is the hourly load at hour $t$. 
\par The annual load profile in Fig. \ref{ANNRES} highlights the effect of seasons on the residential load profile, where it the consumption peaks in summer months starting from May to July and it less in winter months starting from October to February. The effect of weather resulted in an increase in electricity usage of more than 130\% between February and June.
It is observed that the average load is approximately 86\% of the peak load in most months. 
The primary reason for this change is the air conditioning loads as will be investigated in the next section.

\section{Residential Load Structure and Load Composition}
To apply demand side management, a precise specification of the load composition is essential in order to know the contribution percentage of every device to the overall load profile. In the previous section, the daily and annual load profiles were presented using a top-down approach. In this section, a bottom-up approach will be used and matched with the top-down approach. 

\par Data acquired by the Saudi Arabian General Authority for Statistics (GAS) in 2016, shows the usage hours of different home appliances and devices per day \cite{general_authority_for_statistics} (Table \ref{homeactivity}). We rely on this data for the bottom-up approach, where we back-project the consumption of each device/appliance to the top-down load profile to properly scale the percentage of consumption of each load activity. 
\par Considering that the consumption of devices might be affected by the level of loading, we introduce two loading states, i.e., active and idle. The total energy consumption $C$ (\$/kWh) of each load is given as:
\begin{equation}
    C^s = \left(\omega_{R}^{d}\gamma_{R}^{d,s} + \omega_{I}^{d}\gamma_{I}^{d,s}\right)t^{d,s} 
\end{equation}
where $s=\{0,1\}$ is a set resembling winter and summer seasons, respectively. Winter months are assumed to be from October to February and the period from March to September is considered as summer. $t\in \mathbb{Z}_{+}$ is the device daily time of use measured in hours.
$\omega_{R}, \omega_{I}\in \mathbb{R}_{+}$ are the wattage consumption in run-mode and idle-mode, respectively. $\gamma_{R}^{d,s}$ and $\gamma_{I}^{d,s}$ are the fractions of run-time and idle-time of device $d$, respectively, where $d$ is a vector $\bar{d} \in  \mathbb{Z}_{+}^{m}$ s.t. for $m>j>0, d=\{d_{j},d_{j+1},...,d_{m}\}$, where $m$ is the total number of devices or appliances. \{$\gamma_{R},\gamma_{I} \in\mathbb{Q}_+$\} is given such that:
\begin{equation}
     \gamma_{R}+\gamma_{I}=1
\end{equation}

 \par 
Using the information in Table \ref{homeactivity}, the following equation is applied to acquire the average energy consumed by every device:

\begin{equation}
\Psi^{d,s}=q^{d,s} C^{s}
\label{loadeqn}
\end{equation}
 $\Psi^{d,s}$ is the energy consumed by device $d$ in season $s$. Lastly, \{$q\in  \mathbb{N}$\} is the number of units of device $d$ in a household.

\begin{table*}[!htb]
\centering
\caption{Household Activities and Calculation Parameters}
\label{homeactivity}
\resizebox{\textwidth}{!}{%
\begin{tabular}{@{}cccccccccc@{}}
\toprule
Activity & \begin{tabular}[c]{@{}c@{}}ToU\\ (winter)\end{tabular} & \begin{tabular}[c]{@{}c@{}}ToU\\ (summer)\end{tabular} & \begin{tabular}[c]{@{}c@{}}Number of \\ units (winter)\end{tabular} & \begin{tabular}[c]{@{}c@{}}Number of\\ units (summer)\end{tabular} & Rating (W) & \begin{tabular}[c]{@{}c@{}}Idle rating\\ (W)\end{tabular} & Operation & \begin{tabular}[c]{@{}c@{}}Run-time\\  (\%)\end{tabular} & \begin{tabular}[c]{@{}c@{}}Idle-time\\  (\%)\end{tabular} \\ \midrule
\begin{tabular}[c]{@{}c@{}}Heating\\ (oil-filled)\end{tabular} & 8 & 1.5 & 2 & 1 & 1500 & 0 & \begin{tabular}[c]{@{}c@{}}Semi\\ Auto\end{tabular} & 0.5 & 0.5 \\
\begin{tabular}[c]{@{}c@{}}Air\\ conditioning\end{tabular} & 3 & 10 & 2 & 5 & 1800 & 100 & \begin{tabular}[c]{@{}c@{}}Semi\\ Auto\end{tabular} & 0.6 & 0.4 \\
\begin{tabular}[c]{@{}c@{}}Water\\ heating\end{tabular} & 14 & 4.7 & 3 & 1 & 1500 & 30 & Auto & 0.3 & 0.7 \\
\begin{tabular}[c]{@{}c@{}}Water\\ coolers\end{tabular} & 10 & 17 & 1 & 1 & 250 & 10 & Auto & 0.5 & 0.5 \\
\begin{tabular}[c]{@{}c@{}}Water Bump\\ (Dynamo)\end{tabular} & 1.5 & 2.1 & 1 & 1 & 250 & 0 & Auto & 1 & 0 \\
\begin{tabular}[c]{@{}c@{}}Washing \&\\ Drying\end{tabular} & 1.3 & 1.9 & 2 & 2 & 2000 & 0 & \begin{tabular}[c]{@{}c@{}}Semi\\ Auto\end{tabular} & 1 & 0 \\
Ironing & 1 & 1.8 & 1 & 1 & 1000 & 0 & Manual & 1 & 0 \\
\begin{tabular}[c]{@{}c@{}}Vacuum\\ cleaning\end{tabular} & 1 & 1.3 & 1 & 1 & 1000 & 0 & Manual & 1 & 0 \\
Cooking & 1.6 & 1.4 & 1 & 1 & 2150 & 0 & \begin{tabular}[c]{@{}c@{}}Semi\\ Auto\end{tabular} & 1 & 0 \\
\begin{tabular}[c]{@{}c@{}}Electric\\ kettle\end{tabular} & 1.3 & 2 & 1 & 1 & 1800 & 0 & Manual & 1 & 0 \\
Lighting & 7.3 & 7.5 & 50 & 50 & 10 & 0 & Manual & 1 & 0 \\
\begin{tabular}[c]{@{}c@{}}Food\\ preservation\end{tabular} & 24 & 24 & 2 & 2 & 100 & 0 & Auto & 1 & 0 \\
TV & 5.3 & 5.9 & 1 & 2 & 120 & 13 & Manual & 1 & 0 \\
PC & 2.1 & 2.6 & 2 & 2 & 150 & 7.5 & Manual & 1 & 0 \\
\begin{tabular}[c]{@{}c@{}}Gaming\\ devices\end{tabular} & 2.6 & 3 & 4 & 4 & 30 & 7.5 & Manual & 1 & 0 \\ \bottomrule
\end{tabular}%
}
\end{table*}

\section{Results}
\begin{figure}[!htb] 
    \centering
    \includegraphics[trim=2cm 1cm 6.2cm 0.2cm,clip, width=\linewidth,height=8cm]{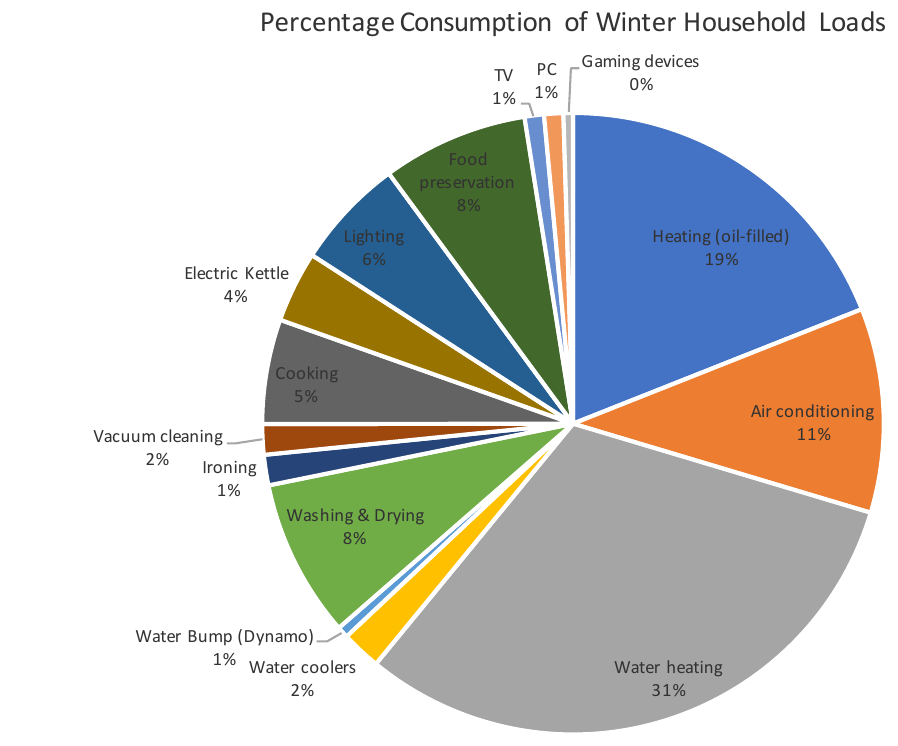} 
    \caption{Percentage of household appliances in winter.} 
    \label{winterL} 
    \end{figure}

\begin{figure}[!htb] 
    \centering
    \includegraphics[trim=2cm 1cm 5.8cm 0.2cm,clip, width=\linewidth,height=8cm]{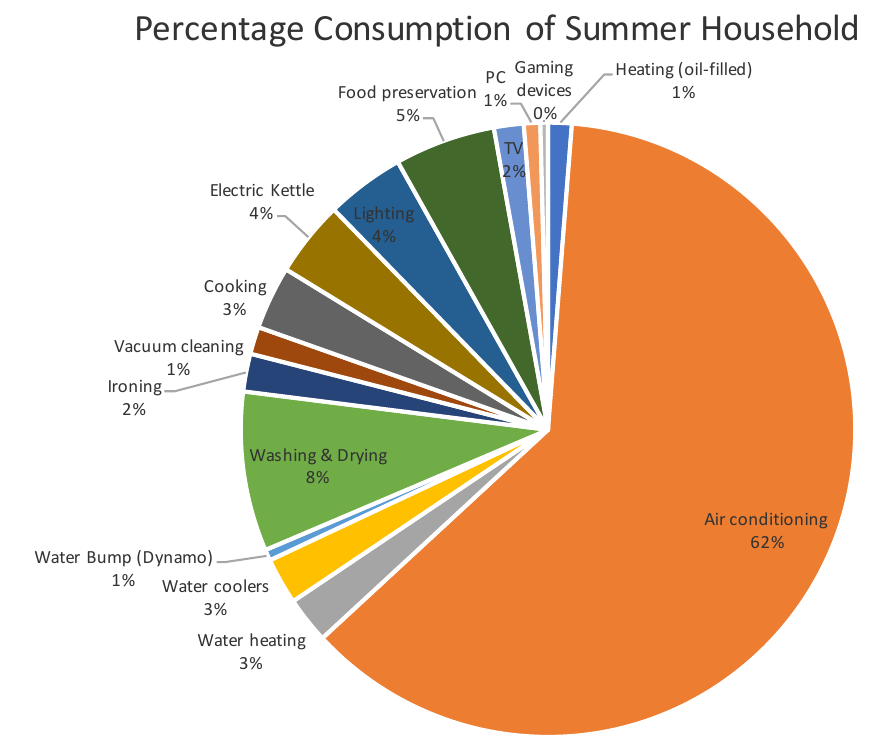} 
    \caption{Percentage of household appliances in summer.} 
    \label{summerL} 
\end{figure}

After applying the collected data on our model in (\ref{loadeqn}), the consumption of each load activity in summer and winter seasons is calculated (Table \ref{AECperactivity}). By matching the residential load profile in Fig.\ref{dailyRES}, with the average energy consumed by every device using (\ref{loadeqn}), the pie-charts in Fig. \ref{winterL} and Fig. \ref{summerL} are acquired. The pie-charts show the percentage share of each home activity during summer and winter months. From Fig. \ref{summerL}, it is observed that Air-conditioning (AC) loads sums up to 62\% of the total load in summer months. This finding is proven as SEC announced a similar percentage in their reports \cite{saudi_electricity_company}. AC loads shrink to only 11\% in winter as the temperature goes down, thus the heating loads rise, and we found that water and ambient heating constitutes to 50\% of the monthly winter demand. As an indicator of the accuracy of the composition percentages, we observe that lighting in winter represented 6\% of the profile, while it was 4\% in summer. The reason is that at winter the evening time is longer and even the day time is cloudy sometimes; therefore, lighting usage in the winter is higher than summer. Other activities such as washing/drying, ironing, and cleaning have very similar consumption in both periods. It is worth mentioning that social behavior and psychological studies can also utilize these findings to investigate the home users patterns and the implication of seasons or holidays.
These findings are the cornerstone of precise and successful application of Direct load control (DLC) by utilities, which in turn helps in maintaining grid and price stability, especially if the control is intelligent enough to be priority-based aware \cite{ALAHMED2020106404} 

\begin{table}[!htb]
\centering
\caption{Summer and winter energy consumption per activity}
\label{AECperactivity}
\resizebox{0.48\textwidth}{!}{%
\centering
\begin{tabular}{@{}ccc@{}}
\toprule
\multicolumn{1}{l}{\centering{Activity}} & \multicolumn{1}{l}{\begin{tabular}[c]{@{}l@{}}Winter Energy \\ Consumption (Wh/day)\end{tabular}} & \multicolumn{1}{l}{\begin{tabular}[c]{@{}l@{}}Summer Energy\\ Consumption (Wh/day)\end{tabular}} \\ \midrule
Heating (oil-filled) & 12000 & 1125 \\
Air conditioning & 6720 & 56000 \\
Water heating & 19782 & 2213.7 \\
Water coolers & 1300 & 2210 \\
Water Bumb (Dynamo) & 375 & 525 \\
Washing \& Drying & 5200 & 7600 \\
Ironing & 1000 & 1800 \\
Vacuum cleaning & 1000 & 1300 \\
Cooking & 3440 & 3010 \\
Electric kettle & 2340 & 3600 \\
Lighting & 3650 & 3750 \\
Food preservation & 4800 & 4800 \\
TV & 636 & 1416 \\
PC & 630 & 780 \\
Gaming devices & 312 & 360 \\
\begin{tabular}[c]{@{}c@{}}Total \\ (KWh/month)\end{tabular} & 1895.55 & 2714.69 \\ \bottomrule
\end{tabular}}%
\end{table}


\section{Conclusion}
In order to control the demand side, detailed knowledge of the load composition and the percentage of each load must be gained. The information about the load composition and percentages is highly correlated with time, season, weather and many other sociological and behavioural factors.
\par This paper presented a simplified approach for finding and understanding load compositions and percentage of each load type. A hybrid top-down and bottom-up approach was used. The top-down approach is represented by acquiring the data from SCADA systems on residential substation. Then the bottom-up approach maps the acquired load composition with the load profile acquired from the top-down approach to obtain the household load composition and the corresponding percentages. The paper uses real and local load data from the eastern province in Saudi Arabia to apply the methodology and present the findings.
\par Using this simple model, the percentage and composition of the household in the region of study were both identified. The acquired data highly matches the anticipated and historical load composition and percentages. 

\section*{Acknowledgment}
The authors would like to acknowledge the support provided by the Deanship of Scientific Research (DSR) at King Fahd University of Petroleum and Minerals (KFUPM) for funding this work through project No. IN161043.

\ifCLASSOPTIONcaptionsoff
  \newpage
\fi

\IEEEtriggeratref{8}
\IEEEtriggercmd{\enlargethispage{-5in}}



%

\printbibliography 
\end{document}